\documentclass[aps,preprint,floats,epsf,epsfig,nofootinbib,letter]{revtex4}
\usepackage{epsfig}

\begin{document}
\def\be{\begin{eqnarray}}
\def\en{\end{eqnarray}}
\def\non{\nonumber}
\def\la{\langle}
\def\ra{\rangle}
\def\nc{N_c^{\rm eff}}
\def\vp{\varepsilon}
\def\drho{\bar\rho}
\def\deta{\bar\eta}
\def\CP{{\it CP}~}
\def\a{{\cal A}}
\def\B{{\cal B}}
\def\c{{\cal C}}
\def\d{{\cal D}}
\def\e{{\cal E}}
\def\p{{\cal P}}
\def\t{{\cal T}}
\def\up{\uparrow}
\def\dw{\downarrow}
\def\vma{{_{V-A}}}
\def\vpa{{_{V+A}}}
\def\smp{{_{S-P}}}
\def\spp{{_{S+P}}}
\def\J{{J/\psi}}
\def\ov{\overline}
\def\Lqcd{{\Lambda_{\rm QCD}}}
\def\pr{{Phys. Rev.}~}
\def\prl{{Phys. Rev. Lett.}~}
\def\pl{{Phys. Lett.}~}
\def\np{{Nucl. Phys.}~}
\def\zp{{Z. Phys.}~}
\def\lsim{ {\ \lower-1.2pt\vbox{\hbox{\rlap{$<$}\lower5pt\vbox{\hbox{$\sim$}
}}}\ } }
\def\gsim{ {\ \lower-1.2pt\vbox{\hbox{\rlap{$>$}\lower5pt\vbox{\hbox{$\sim$}
}}}\ } }

\font\el=cmbx10 scaled \magstep2{\obeylines\hfill March, 2008}

\vskip 1.5 cm

\centerline{\large\bf Baryonic $D$ Decay $D_s^+\to p\bar n$ and Its Implication}

\bigskip
\centerline{\bf Chuan-Hung Chen,$^1$ Hai-Yang Cheng$^2$, and Yu-Kuo Hsiao$^2$}
\medskip
\medskip
\centerline{$^1$ Physics Department, National Cheng-Kung University}
\centerline{Tainan, Taiwan 700, Republic of China}
\medskip
\medskip
\centerline{$^2$ Institute of Physics, Academia Sinica}
\centerline{Taipei, Taiwan 115, Republic of China}

\medskip
\bigskip
\bigskip
\centerline{\bf Abstract}
\bigskip

\small The channel $D_s^+\to p\bar n$ is the only kinematically allowed baryonic $D$ decay. It proceeds solely through the $W$-annihilation topology. Hence, a recent observation of this mode by CLEO will shed light on the dynamics of $W$-annihilation.
At the short-distance level, its branching ratio is very small, of order $10^{-6}$, owing to chiral suppression. It receives long-distance contributions through final-state scattering of the leading tree and color-suppressed amplitudes. Assuming that the long-distance enhancement of $W$-annihilation in the baryonic $D$ decay is similar to that in the mesonic $D_s^+$ decay, where the latter can be obtained from the analysis of the diagrammatic approach, we find that $D_s^+\to p\bar n$ becomes visible. The observation of this baryonic $D$ decay implies the dynamical enhancement of the $W$-annihilation topology in the $D_s^+$ decay.

\pagebreak

\small

1. It is well known that the $B$ meson is
heavy enough to allow a baryon-antibaryon pair production in the
final state. In the past two decades, there has been much progress in the study of baryonic $B$ decays, both experimentally and theoretically (for a review, see \cite{Cheng05}).
However, it is less known that the baryonic decay can also occur in the charm sector. It turns out that there is only one
baryonic $D$ decay mode which is physically allowed, namely, $D_s^+\to p\bar n$. Since this decay mode can only proceed through $W$-annihilation at the short-distance level,
its branching ratio is expected to be very small, of order
$10^{-6}$, because of partial conservation of axial current
\cite{Pham1}. That is, the decay amplitude is subject to chiral suppression. As pointed out by Pham long time ago \cite{Pham2}, if the decay width of $D_s^+$ is dominated by $W$-annihilation, the aforementioned chiral suppression will be alleviated and its branching ratio could be enhanced to the percent level. Very recently, this mode was first observed by CLEO with the result $\B(D_s^+\to p\bar n)=(1.30\pm0.36^{+0.12}_{-0.16})\times 10^{-3}$ \cite{CLEO}. Recall that none of the charmless two-body baryonic $B$ decays e.g. $B\to p\bar p,\Lambda\bar\Lambda, \Lambda\bar p$ has been observed and the experimental limit on branching ratios has been pushed to the level of $10^{-7}$  \cite{Belle}.

The first observation of the baryonic $D$ decay $D_s^+\to p\bar n$ will shed light on the importance of the $W$-annihilation mechanism in charm decays. In this letter, we will discuss its implication.

\vskip 0.5cm
2. At the short-distance level, the baryonic decay $D_s^+\to p\bar n$ proceeds through $W$-annihilation as depicted in Fig. 1. It is known that the short-distance $W$-annihilation is very small as it is subject to helicity suppression.
In the factorization approach, its amplitude has the expression
\be
A(D_s^+\to p\bar n)={G_F\over\sqrt{2}}V_{cs}V_{ud}^*\,a_1f_{D_s}q^\mu\la p\bar n|\bar u\gamma_\mu(1-\gamma_5)d|0\ra,
\en
where $q=p_P+p_{\bar n}$, $a_1$ of order unity is an effective Wilson coefficient and $f_{D_s}$ is the decay constant of the $D_s^+$ meson. The $p\bar n$ matrix element can be expressed in terms of six form factors
\be
\la p(p_p)\bar n(p_{\bar n})|(V-A)_\mu|0\ra &=& \bar
u_p(p_p)\Bigg\{f_1^{p\bar n}(q^2)\gamma_\mu+i{f_2^{p\bar n}(q^2)\over
m_p+m_n} \sigma_{\mu\nu}q^\nu+{f_3^{p\bar n}(q^2)\over
m_p+m_n}q_\mu \non  \\ && -
\Big[g_1^{p\bar n}(q^2)\gamma_\mu+i{g_2^{p\bar n}(q^2)\over
m_p+m_n} \sigma_{\mu\nu}q^\nu+{g_3^{p\bar n}(q^2)\over
m_p+m_n}q_\mu\Big]\gamma_5\Bigg\}v_n(p_{\bar n}).
\en
Applying equations of motion to the quark currents
\be
 -i\partial^\mu(\bar q_1\gamma_\mu q_2)=(m_1-m_2)\bar q_1 q_2,
 \qquad\quad -i\partial^\mu(\bar q_1\gamma_\mu\gamma_5 q_2)=(m_1+m_2)\bar q_1\gamma_5
 q_2,
\en
it is clear that only the axial-vector current will contribute to the dibaryon matrix element and the decay amplitude vanishes in the chiral limit. Hence, helicity suppression is manifested in the factorization approach as CVC and PCAC.
Neglecting the tiny isospin violation, we obtain
\be
A(D_s^+\to p\bar n)={G_F\over\sqrt{2}}V_{cs}V_{ud}^*\,a_1f_{D_s}\left(2m_Ng_1^{p\bar n}+{t\over 2m_N}g_3^{p\bar n}\right)\bar u_p\gamma_5 v_{\bar n},
\en
where $t=q^2=m_{D_s}^2$.
Since the pseudoscalar form factor $g_3^{p\bar n}$
corresponds to a pion pole contribution to the $p\bar n$
axial matrix element, it follows that
 \be
 g_3^{p\bar n}(t)=-{4m_N^2\over t-m_\pi^2}g_1^{p\bar n}(t).
 \en
Consequently,
\be
A(D_s^+\to p\bar n)={G_F\over\sqrt{2}}V_{cs}V_{ud}^*\,a_1f_{D_s}2m_N\,\left({m_\pi\over m_{D_s}}\right)^2g_1^{p\bar n}(m_{D_s}^2)\bar u_p\gamma_5 v_{\bar n}.
\en
The chiral suppression factor of $m_\pi^2/m_{D_s}^2$ follows from the PCAC relation, as it should.

There is not much information on the form factor $g_1^{p\bar n}$ at $q^2=m_{D_s}^2$. At $q^2=0$ we have $g_1^{p\bar n}(0)=-1.27$. At large $q^2$,
we can reply on pQCD to consider its asymptotic behavior \cite{Brodsky}
 \be \label{eq:larget}
 g^{p\bar n}_1(t) \to {5\over 3}G_M^p(t)+G_M^n(t),
 \en
where $G_M^{p,n}$ are the nucleon's magnetic form factors. A  phenomenological fit to the experimental data of nucleon's electromagnetic
form factors is available in \cite{CHT}  using the
following parametrization:
 \be \label{GMN}
 |G_M^p(t)| &=& \left({x_1\over t^2}+{x_2\over t^3}+{x_3\over t^4}
 +{x_4\over t^5}+{x_5\over t^6}\right)\left[\ln{t\over
 Q_0^2}\right]^{-\gamma},  \non \\
|G_M^n(t)| &=& \left({y_1\over t^2}+{y_2\over
t^3}\right)\left[\ln{t\over Q_0^2}\right]^{-\gamma},
 \en
where $Q_0=\Lambda_{\rm QCD}$  and $\gamma=2+{4\over
3\beta}=2.148$\,. Following the best fit obtained in \cite{CHT}, we find $g_1^{p\bar n}(m_{D_s}^2)\approx -0.22$. Since the relation (\ref{eq:larget}) holds in the $t\to\infty$ limit, we will allow $g_1^{p\bar n}(m_{D_s}^2)$ to be varied by a factor of 2.

\begin{figure}[t]
\vspace{0cm}
\centerline{\epsfig{figure=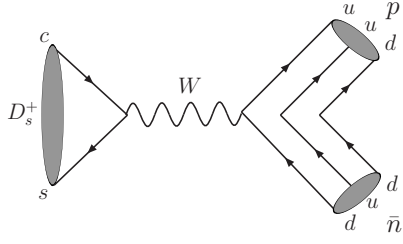,width=6cm}}
    \caption[]{\small Quark diagram for $D_s^+\to p\bar n$.}
    \label{fig:f1K}
\end{figure}

For the general baryonic decay amplitude given by
 \be
 M(D\to \B_1\ov \B_2)=\bar u_1(A+B\gamma_5)v_2,
 \en
with $A$ and $B$ corresponding to $p$-wave parity-violating and
$s$-wave parity-conserving amplitudes, respectively, the
decay rate reads
\begin{eqnarray}
\Gamma(D\to {\cal B}_{1}(1/2^{+}) \bar {\cal B}_{2}(1/2^{+}))&=&
\frac{p_{c}}{4\pi m^2_{D}} \left\{ |A|^2 \left(m^2_{D} -(m_2 +
m_1)^2 \right) \right. \nonumber \\
&+&\left. |B|^2  \left(m^2_{D} -(m_2 - m_1)^2 \right)\right\},
\end{eqnarray}
where $p_c$ is the c.m. momentum and $m_i$ is
the mass of the baryon $\B_i$. Putting everything together, we obtain
\be
\B(D_s^+\to p\bar n)_{\rm SD}=(0.4^{+1.1}_{-0.3})\times 10^{-6},
\en
where use of $f_{D_s}=282$ MeV has been made.
The theoretical error is due to the uncertainty in the form factor $g_1^{p\bar n}(m_{D_s}^2)$.

\vskip 0.5cm
3. Although the short-distance weak annihilation contributions,
namely, $W$-exchange and $W$-annihilation, are small and negligible based on the helicity suppression argument, it was realized in 1980s that the long-distance contribution to weak annihilation in charm decays
can be sizable. For example,
the observation of $D^0\to\ov K^0\phi$ in the middle 1980s gave
the first clean evidence of $W$-exchange. Hence, the alleviation of the helicity suppression on $W$-annihilation may render the decay $D_s^+\to p\bar n$ detectable.

It has been established that a least
model-independent analysis of heavy meson decays can be carried
out in the so-called quark-diagram approach \cite{Chau,CC86,CC87}. In the diagrammatic approach, all two-body nonleptonic weak decays
of heavy mesons can be expressed in terms of six distinct quark
diagrams :\footnote{Historically, the quark-graph
amplitudes $\t,\,\c,\,\e,\,\a$ were originally denoted by
$\a,\,\B,\,\c,\,\d$, respectively \cite{Chau,CC86,CC87}.}
$\t$, the color-allowed external
$W$-emission tree diagram; $\c$, the color-suppressed internal
$W$-emission diagram; $\e$, the $W$-exchange diagram; $\a$, the
$W$-annihilation diagram; ${\cal P}$, the penguin diagram; and
${\cal V}$, the vertical $W$-loop diagram. It should be stressed
that these quark diagrams are classified according to the
topologies of weak interactions with all strong interaction
effects included and hence they are {\it not} Feynman graphs. All
quark graphs used in this approach are topological with all the
strong interactions included, i.e. gluon lines are included in all
possible ways.

As stressed above, topological graphs can provide information on
final-state interactions (FSIs). In general, there are several
different forms of FSIs: elastic scattering and inelastic
scattering such as quark exchange, resonance formation,$\cdots$,
etc. \footnote{The effects of the nearby resonances on
weak annihilation in charm decays have been discussed in \cite{Chenga1a2}.}
Take the decay $D_s^+\to p\bar n$ as an illustration. The
topological amplitude $\a$ can receive contributions from final-state rescattering of
the tree amplitude $\t$ of e.g. $D_s^+\to \pi^+\eta^{(')}$ and the color-suppressed amplitude $\c$ of $D_s^+\to K^+\bar K^0$ (see Fig. 2). They have the same topology as $W$-annihilation. Since these mesonic $D_s^+$ decays have branching ratios of order $10^{-2}$, more precisely \cite{CLEODs}, \footnote{The new CLEO results \cite{CLEODs} are smaller than the branching fractions
$\B(D_s^+\to\pi^+\eta')=(4.7\pm0.7)\%$, $\B(D_s^+\to\pi^+\eta)=(2.11\pm0.35)\%$ and $\B(D_s^+\to K^+\bar K^0)=(4.4\pm0.9)\%$ cited in the Particle Data Group \cite{PDG}.}
\be \label{eq:CLEODs}
&& \B(D_s^+\to\pi^+\eta')=(3.77\pm0.39)\%, \qquad \B(D_s^+\to\pi^+\eta)=(1.58\pm0.21)\%, \non \\ && \B(D_s^+\to K^+\bar K^0)=(2.98\pm0.17)\%,
\en
it is thus conceivable that $\B(D^+_s\to p\bar n)$ induced from final-state rescattering can reach the level of $10^{-3}$.
 Therefore, {\it even if the
short-distance $W$-annihilation vanishes, a long-distance $W$-annihilation
can be induced via final-state rescattering}. Historically, it was first
pointed out in \cite{Donoghue} that rescattering effects required
by unitarity can produce the reaction $D^0\to\ov K^0\phi$, for
example, even in the absence of the $W$-exchange diagram. Then it
was shown in \cite{CC87} that this rescattering diagram belongs to
the generic $W$-exchange topology.

Contrary to the $B$ decays,
the charmed meson is not heavy enough to allow for  a sensible approach based on the heavy quark expansion, such as QCD factorization \cite{BBNS}, pQCD \cite{pQCD} and soft-collinear effective theory \cite{SCET}. Nevertheless, it has some unique advantages over $B$ physics, namely, many of the topological amplitudes, especially $W$-exchange and $W$-annihilation, can be extracted from the data.
Various diagrammatic
amplitudes have been inferred from the measured two-body $D$ decays in \cite{Rosner,Zhong}.
One of the important observations one can learn from these
analyses is that the weak annihilation ($W$-exchange or
$W$-annihilation) amplitude is sizable with a large phase relative
to the tree amplitude.

\begin{figure}[t]
\vspace{0cm}
\centerline{\epsfig{figure=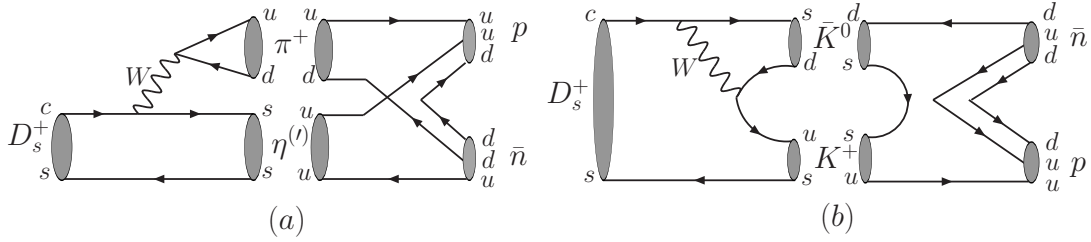,width=15cm}}
    \caption[]{\small Long-distance contributions to $D_s^+\to p\bar n$ via final-state rescattering of (a) the $W$-emission amplitude of $D_s^+\to\pi^+\eta^{(')}$ and (b) the color-suppressed amplitude of $D_s^+\to K^+\bar K^0$. Both diagrams have the same topology as $W$-annihilation.}
    \label{fig:DSI}
\end{figure}

Although we understand qualitatively the enhancement of $W$-annihilation via final-state rescattering, it is difficult to make a quantitative statement about FSI effects in Fig. 2. \footnote{In principle, final-state rescattering effects can be phenomenologically  modeled as one-particle-exchange processes at the hadron level (see e.g. \cite{CCSfsi}). However, this task will be much more difficult for the baryonic decays.}
Nevertheless, it is plausible to assume that the enhancement of $W$-annihilation in the baryonic $D$ decay is similar to that in the mesonic  decay $D_s^+\to \pi^+\eta_q$; that is,
\be \label{eq:ratio}
 {A(D_s^+\to p\bar n)\over A(D_s^+\to p\bar n)_{\rm SD}}\approx
 {A(D_s^+\to \pi^+\eta_q)\over A(D_s^+\to \pi^+\eta_q)_{\rm SD}},
 \en
where $\eta_q$ and $\eta_s$ are defined as
 \be
 \eta_q={1\over\sqrt{2}}(u\bar u+d\bar d),\qquad\quad
 \eta_s=s\bar s,
 \en
in analog to the wave functions of $\omega$ and $\phi$ in ideal
mixing. The wave functions of the $\eta$ and $\eta'$ are given by
 \be \label{eq:etawf}
 \left(\matrix{ \eta \cr \eta'\cr}\right)=\left(\matrix{ \cos\phi & -\sin\phi \cr
 \sin\phi & \cos\phi\cr}\right)\left(\matrix{\eta_q \cr \eta_s
 \cr}\right).
 \en
In terms of the topological diagrams,
\be
A(D_s^+\to K^+\bar K^0)=\c+\a,\qquad  A(D_s^+\to \pi^+\eta_q)=\sqrt{2}\a,\qquad  A(D_s^+\to \pi^+\eta_s)=\t.
\en

A simple calculation based on factorization yields
\be \label{eq:Dsamp}
A(D_s^+\to\pi^+\eta_q)_{\rm SD} &=& 2\,{G_F\over\sqrt{2}}V_{cs}V_{ud}^*\,a_1f_{D_s}(m_{\eta_q}^2-m_\pi^2)F_0^{\pi\eta_q}(m_{D_s}^2), \non \\ A(D_s^+\to\pi^+\eta_s)_{\rm SD} &=& {G_F\over\sqrt{2}}V_{cs}V_{ud}^*\,a_1f_\pi(m_{D_s}^2-m_{\eta_s}^2)F_0^{D_s\eta_s}(m_\pi^2).
\en
Contrary to $D_s^+\to p\bar n$, only the vector current will contribute to the $\pi\eta_q$ matrix element in the decay $D_s^+\to\pi^+\eta_q$.
Since the short-distance $W$-annihilation vanishes in the chiral limit, the form factor $F_0^{\pi\eta_q}(q^2)$ is expected to be of order $m_\pi\Lambda_{\rm QCD}/q^2$ . The masses of $\eta_q$ and $\eta_s$ read \cite{Feldmann}
\be
m_{\eta_q}^2 &=& {\sqrt{2}\over f_q}\la 0|m_u\bar ui\gamma_5u+m_d\bar di\gamma_5d|\eta_q\ra+{\sqrt{2}\over f_q}\la 0|{\alpha_s\over 4\pi}G\tilde G|\eta_q\ra\approx m_\pi^2+ {\sqrt{2}\over f_q}\la 0|{\alpha_s\over 4\pi}G\tilde G|\eta_q\ra \non \\
m_{\eta_s}^2 &=& {2\over f_s}\la 0|m_s\bar si\gamma_5s|\eta_s\ra+{1\over f_s}\la 0|{\alpha_s\over 4\pi}G\tilde G|\eta_s\ra\approx 2m_K^2-m_\pi^2+ {1\over f_s}\la 0|{\alpha_s\over 4\pi}G\tilde G|\eta_s\ra,
\en
where $f_q$, $f_s$ are the decay constants of $\eta_q$ and $\eta_s$, respectively, and contributions to their masses from the gluonic anomaly have been included. We shall use the parameters extracted from a phenomenological fit \cite{Feldmann}: $\phi=(39.3\pm1.0)^\circ$ and
\be
a^2 &\equiv & {1\over \sqrt{2}f_q}\la 0|{\alpha_s\over 4\pi}G\tilde G|\eta_q\ra=0.265\pm0.010, \non \\
y &\equiv &{\sqrt{2}\la 0|{\alpha_s\over 4\pi}G\tilde G|\eta_s\ra \over
\la 0|{\alpha_s\over 4\pi}G\tilde G|\eta_q\ra}={f_q\over f_s}=0.81\pm0.03\,.
\en

Since a fit to the data  (\ref{eq:CLEODs})
cannot fix the magnitude of $\t$ and $\a$ and their relative phase simultaneously, we can reply on either the factorization calculation for $\t$ using $F_0^{D_s\eta_s}(0)=0.78$ \cite{Melikhov}
or the diagrammatic amplitudes inferred from a global fit to Cabibbo-allowed $D\to PP$ data in conjunction with SU(3) symmetry.
The former leads to $\t\approx 2.6\times 10^{-6}$ GeV, which is very close to $\t\approx 2.7\times 10^{-6}$ GeV obtained in \cite{Rosner}. For convenience we take the tree amplitude $\t$ to be real. We find that a fit to the data of $D_s^+\to \pi\eta$ and $\pi\eta'$ yields  $\a_{\rm exp}\approx 0.68\exp(- i55^\circ)\times 10^{-6}$ GeV, where the sign of the phase is fixed by the $D_s^+\to K^+\bar K^0$ rate.\footnote{Our result differs slightly from the one $\a_{\rm exp}= (0.54\pm0.37)\exp[-i(64^{+32}_{-~8})^\circ]\times 10^{-6}$ GeV quoted in \cite{CLEODsSU3} since we use the realistic angle $\approx 39.3^\circ$ for the  $\eta-\eta'$ mixing rather than the ``magic" one $\phi=35.2^\circ$ as employed in \cite{CLEODsSU3}. Note that Eq. (\ref{eq:etawf}) is simplified to $\eta=(\sqrt{2}\eta_q-\eta_s)/\sqrt{3}$ and $\eta'=(\eta_q+\sqrt{2}\eta_s)/\sqrt{3}$ for the latter mixing angle.}
Putting this back to Eq. (\ref{eq:ratio}) leads to
 \be
 \B(D_s^+\to p\bar n)\approx  \left(0.8^{+2.4}_{-0.6}\right)\times 10^{-3},
 \en
 where use of $\Lambda_{\rm QCD}\approx 250$ MeV has been made and
 only the theoretical uncertainties due to the form factor $g_1^{p\bar n}(m_{D_s}^2)$ have been taken into account. The result is consistent with the CLEO measurement $\B(D_s^+\to p\bar n)=(1.30\pm0.36^{+0.12}_{-0.16})\times 10^{-3}$ \cite{CLEO}.
 Therefore, the above crude estimate suffices to demonstrate that the branching fraction of $D^+\to p\bar n$ can be easily enhanced to the 0.1\% level by the long-distance enhancement to $W$-annihilation.

\vskip 0.5cm
4. In short, the decay $D_s^+\to p\bar n$  proceeds solely through the $W$-annihilation topology and is the only baryonic $D$ decay that is physically allowed.
Hence, a recent observation of this mode by CLEO will shed light on the dynamics of $W$-annihilation.
At the short-distance level, its branching ratio is very small, of order $10^{-6}$, owing to chiral suppression. It receives long-distance contributions through final-state scattering of the leading tree and color-suppressed amplitudes. Assuming that the long-distance enhancement of $W$-annihilation in the baryonic $D$ decay is similar to that in the mesonic $D_s^+$ decay, where the latter can be obtained from the analysis of the diagrammatic approach, we find that $D_s^+\to p\bar n$ becomes visible. The observation of this baryonic $D$ decay implies the dynamical enhancement of the $W$-annihilation topology in the $D_s^+$ decay.

Finally, we would like to remark that the baryonic  decay $D_s^+\to p\bar n$ should be readily accessible to BESIII. Therefore, a confirmation of this unique mode by BESIII will be highly desirable.

\vskip 2.3cm \acknowledgments One of us (HYC) wishes to thank Ikaros Bigi for bringing reference \cite{Pham2} to his attention three years ago.
 This research was supported in part by the National
Science Council of R.O.C. under Grant Nos. NSC96-2112-M-001-003 and NSC96-2112-M-013-MY2.

\vskip 1.0cm {\it Note Added}: After the paper was submitted for publication, we learned that this baryonic $D$ decay has been considered by I. Bediaga and E. Predazzi, Phys. Lett. B {\bf 275}, 161 (1992).


\end{document}